\documentclass{article}
\usepackage{maple2e}
\usepackage{amsmath}
\DefineParaStyle{Maple Output}
\DefineCharStyle{2D Math}
\DefineCharStyle{2D Output}
\begin{document}

\title{Invariants of the Riemann tensor for Class B Warped Product
Spacetimes}
\author{Kevin Santosuosso\footnote{Department of Physics, Queen's University, Kingston, Ontario, Canada K7L 3N6} ,
Denis Pollney\footnote{Faculty of Mathematical Studies, University of Southampton, Highfield, Southampton SO17 1BJ UK} \footnote{dp@maths.soton.ac.uk} ,
Nicos Pelavas\footnote{Department of Physics,Queen's University, Kingston, Ontario, Canada K7L 3N6} ,
Peter Musgrave\footnote{West End Systems Corporation,39 Winner's Circle, Arnprior, Ontario, Canada K7S 3G9 }\\ and
Kayll Lake\footnote{Department of Physics,Queen's University,Kingston,Ontario,Canada K7L 3N6}\footnote{lake@astro.queensu.ca}
}

\maketitle

\begin{abstract}
We use the computer algebra system \textit{GRTensorII} to examine invariants
polynomial in the Riemann tensor for class $B$ warped product spacetimes -
those which can be
decomposed into the coupled product of two 2-dimensional spaces, one
Lorentzian and one Riemannian, subject to the separability of the
coupling:
\begin{equation}
ds^2 = ds_{\Sigma_1}^2 (u,v) + C(x^\gamma)^2 ds_{\Sigma_2}^2
(\theta,\phi)
\end{equation}
with $C(x^\gamma)^2=r(u,v)^2 w(\theta,\phi)^2$ and $sig(\Sigma_1)=0$,
$sig(\Sigma_2)=2\epsilon$ ( $\epsilon=\pm 1$) for class $B_{1}$
spacetimes and $sig(\Sigma_1)=2\epsilon$, $sig(\Sigma_2)=0$ for class
$B_{2}$.  Although very special, these spaces include
many of interest, for example, \emph{all} spherical, plane, and
hyperbolic spacetimes.  The first two Ricci invariants along with the
Ricci scalar and the real component of the second Weyl invariant $J$
alone are shown to constitute the largest independent set  of invariants to
degree five for this class. Explicit syzygies are given for other invariants up to this
degree.  It is argued that this set constitutes the largest
functionally independent set to \emph{any} degree for this class, and
some physical consequences of the syzygies are explored.
\end{abstract}

\section{Introduction}
It is natural to suspect that invariants constructed from the Riemann
tensor via contractions\footnote{We do not consider covariant
derivatives of the Riemann tensor.  That is, in the terminology of
\cite{ES}, we consider invariants of order $k=2$.} should play some
important role in general relativity.  However, it is only fair to say
that this importance has yet to be fully realized.  For example, it
has been known for a very long time that this type of invariant fails
to distinguish inequivalent spacetimes \cite{NK}. Indeed, the
inclusion of derivatives of the Riemann tensor does not overcome this
shortcoming \cite{Koutras}.  The inequivalence problem requires a more
sophisticated approach (\textit{e.g.} \cite{McS}).  Invariants are of some use
in the study of spacetime singularities (distinguishing the scalar
polynomial (\emph{sp}) type (\textit{e.g.} \cite{TCE}) which is equivalent to a
curvature \emph{pp} type \cite{Siklos}), but even here they fail to
distinguish shell crossings from more significant singularities.

Despite these shortcomings, the problem of finding a complete set of
invariants for spacetime has attracted considerable recent effort
\cite{CM,Sneddon,ZM,Pollney96}.  Here we set out to study spacetimes
which are the product of two 2-dimensional spaces, one Lorentzian and
one Riemannian, subject to a separability condition on the function
which couples the 2-spaces.  That is, we consider metrics of the form
\begin{equation}
ds^2 = ds_{\Sigma_1}^2 (u,v) + C(x^\gamma)^2 ds_{\Sigma_2}^2
(\theta,\phi)
\end{equation}
subject to the restriction
\begin{equation}
C(x^\gamma)^2=r(u,v)^2 w(\theta,\phi)^2.
\end{equation}
The spaces are known as warped products of class B \cite{CC}, with
$sig(\Sigma_1)=0$, $sig(\Sigma_2)=2\epsilon$ ( $\epsilon=\pm 1$) for
class $B_{1}$ and $sig(\Sigma_1)=2\epsilon$, $sig(\Sigma_2)=0$ for
class $B_{2}$.  A metric of sufficient generality for class $B_{1}$ is
given by\footnote{We assume that all functions are twice continuously
differentiable. We do not consider surface layers. Note that we have
chosen to write $\Sigma_{2}$ in conformally flat form for class
$B_{1}$.} (\textit{e.g.} \cite{Nakahara})
\begin{equation}
ds^2 = -2f(u,v)dudv+r(u,v)^2 g(\theta,\phi)^2 (d\theta^2 + d\phi^2)
\end{equation}
and for class $B_{2}$ by
\begin{equation}
ds^2 = f(u,v)^2(du^2+dv^2)-2r(u,v)^2 g(\theta,\phi) d\theta d\phi.
\end{equation}
There are many examples of class $\rm{B_1}$ spacetimes (for example,
\emph{all} spherical, plane, and hyperbolic spacetimes).  Examples of
spacetimes of class  $\rm{B_2}$ are considerably harder to find.  It
is known that the only physically interesting energy-momentum types
of class $\rm{B_2}$ spacetimes are non-null  electromagnetic,
$\Lambda$-term, or vacuum \cite{HC96}.

\section{The number of invariants required}

For a general Petrov type D metric, we can align the tetrad along the
principle null directions of the Weyl tensor to find a frame in which
there are in general 12 non-zero components of the curvature (complex
$\Psi_2$, the Ricci spinor components, and the Ricci scalar). Since
this alignment leaves a two parameter group of rotational freedom in
the frame (`spin' and `boost'), the number of independent invariants
for a type D spacetime is reduced by this dimension to 10.

Specializing to the given class of spacetimes, we find that if the
above frame alignment is carried out, then a number of the Ricci
spinor components are also reduced to zero. For class $B_1$ we are
left with the real components $\Phi_{00}$, $\Phi_{11}$, and
$\Phi_{22}$, while for class $B_2$, only $\Phi_{11}$ and
$\Phi_{02}=\overline{\Phi}_{20}$ remain non-zero. The number of independent
degrees of freedom arising from the Ricci components is thus reduced
to 3. However, one of these degrees can be removed by making use of
either the spin (for $B_1$) or boost (for $B_2$) freedoms still
remaining in the frame.

The number of independent invariants that we should expect to find
for the given class of spacetime is thus
\begin{displaymath}
   2 + 2 + 1 - 1 = 4,
\end{displaymath}
corresponding respectively to the Weyl and Ricci spinor freedoms, the
Ricci scalar, and the dimension of the invariance group of these
spinors.

An example of how the details of this calculation can be carried out
using \emph{GRTensorII} is given in Appendix A.

\section{The invariants}
The invariants we begin with here are the sixteen suggested by
Carminati and McLenaghan \cite{CM} augmented with an invariant
equivalent to $M$ suggested recently by Zakhary and McIntosh
\cite{ZM}.\footnote{As an aside we note here that if we consider, in the usual way (\textit{e.g.} \cite{HS}),
the secular equation $|S_{ab}-\lambda g_{ab}|=0$, where $S_{ab}$ is
the trace-free Ricci tensor, and $g_{ab}$ is the metric tensor, the
characteristic polynomial which follows is
$\alpha_1\lambda^4+\alpha_2\lambda^3+
\alpha_3\lambda^2+\alpha_4\lambda+\alpha_5=0$.  Since the
$\alpha_{i}^{'s}$ are derived from the components of $S_{ab}$ and
$g_{ab}$ by algebraic operations, they can be called `algebraic'
invariants of the tensor $S_{ab}$.  It can be shown that for this
class of spacetimes the following relationships hold between the
invariants $r_1, r_2, r_3$ and $\alpha_1, \alpha_3, \alpha_4,
\alpha_5$ ($\alpha_2=0$ is a consequence of $S_{ab}$ being
trace-free):
\begin{equation}
\mbox{$r_1=-\frac{\alpha_3}{2\alpha_1},\;\;\;r_2=\frac{3\alpha_4}{8\alpha_1},\;\;\;r_3=\frac{\alpha_3^2}{8\alpha_1^2}
-\frac{\alpha_5}{4\alpha_1}$}.
\end{equation}}  
We adopt the definitions in \cite{CM} for the set\footnote{Two scalars found frequently in the physical literature are the Kretschmann scalar ($\emph{RiemSq} \equiv R_{a b c d}R^{a b c d}$, $R_{a b c d}$ the Riemann tensor) and the square of the Ricci tensor ($\emph{RicciSq} \equiv R_{a b}R^{a b}$, $R_{a b}$ the Ricci tensor). In terms of the set (\ref{CM}) these are given simply by $\emph{RiemSq}=R^{2}/6+8(r_1+Re(w_1))$ and $\emph{RicciSq}=(R^{2}+r_1)/4$.}
\begin{equation}
CM = \{R,r_1,r_2,r_3,w_1,w_2,m_1,m_2,m_3,m_4,m_5\} \label{CM}
\end{equation}
and augment these with
\begin{equation}
ZM = m_6 = \Psi_{ABCD} \Phi^{AE\dot{A}\dot{B}} {\Phi^B}_{E\dot{C}\dot{D}}
  {\Phi^C}_{F\dot{A}\dot{B}} \Phi^{DF\dot{C}\dot{D}} \label{M}.
\end{equation}
  In total this set constitutes six real and six complex invariants.
The completeness of this set has been examined recently \cite{ZM}. It
is certainly complete to degree five (degree being the total number of
Weyl, conjugate Weyl, and Ricci spinors whose contractions make up an
invariant)\footnote{It is a simple (though tedious, unless carried out
electronically) matter to write down expressions for all index
contractions among any set of $n$ curvature spinors. If this is done
to degree five, it is found that for \textit{general} spacetimes every
invariant which is not included in the CM+ZM set satisfies an identity
allowing it to be written as a rational function of members of the
set. Thus, there are no invariants of degree less than or equal to
five to be added to the CM+ZM set. \cite{Pollney96}}. In what follows
we {\it explicitly} reduce this set of eighteen invariants to four for
the spacetimes under consideration by way of
syzygies \cite{PR2}.  All
syzygies have been developed and tested with the aid of the computer
algebra package \emph{GRTensorII} \cite{GRT}.\footnote {GRTensorII is a
package which runs within MapleV. It is entirely distinct from packages
distributed with MapleV and must be obtained independently. The
GRTensorII software and documentation is distributed freely on the
World-Wide-Web from the address {\tt
http://www.astro.queensu.ca/\~{}grtensor/GRHome.html} or {\tt
http://www.maths.soton.ac.uk/\~{}dp/grtensor/}.  Worksheets which
reproduce the syzygies reported here can be downloaded from these
sites.}

\section{Syzygies}
Syzygies follow from a knowledge of which spinor components are non-zero. 
In the present case the important features are that these components 
are real and that the Weyl and Plebanski tensors (\textit{e.g.} \cite{PR2})
share the same principal null directions and are of type $D$ (or $O$).
Since spacetimes of the form (2) with (3) are of Petrov type $D$ (or $O$), 
we have the well-known elementary syzygy\footnote{All syzygies can in fact be obtained \textit{algorithmically}.
By appropriately defining the ``degree" of a syzygy, the most general
syzygy can be constructed for any particular ``degree".  Expanding and 
collecting like terms allows the coefficients to form a linear system 
of equations which are then solved to find the coefficients of the 
syzygy.  If there are no nonzero solutions after the linear system is 
solved, then the syzygy must be of a different ``degree".  The definition
of ``degree" in this case imposes an ordering, allowing a sequential search
to be executed. In any particular case
(\textit{e.g.} as considered in this paper) the number of degrees of freedom can be
counted and so the search for syzygies can be made exhaustive. 
Whereas degree is commonly used as a single parameter, as given in Section 3 above, a
multi-parameter definition of degree 
is more convenient here. For example, an invariant  
can be given a ``degree" $[A,B,C]$, where $A$ refers to the exponent on the Weyl 
spinor, $B$ refers to the number of multiplications in the Ricci spinor, and $C$ 
refers to the exponent on the conjugate Weyl spinor.  
In this notation, for the types of spacetimes we consider here, only two parameters
are necessary, $[A,B]$.  Given a
rule of composition: for invariants $I1=[A1,B1]$ and $I2=[A2,B2]$,
the degree of $I1*I2$ is   
\begin{equation}
                 [A1,B1]+[A2,B2]=[A1+A2,B1+B2],
\end{equation}
the parameter space given by $[A,B]$ can then be searched exhaustively for the 
existence of syzygies.}
\begin{equation}
6w_2^2-w_1^3=0 \label{elm}.
\end{equation}
Furthermore, for any Petrov type $D$ (or $O$) spacetime we have
\cite{Pollney95}
\begin{equation}
(3m_2 - w_1r_1)w_1 -3m_1 w_2 = 0 \label{m2}
\end{equation}
and
\begin{equation}
(3m_5 - w_1 \overline{m}_1)w_1 -3 m_3 w_2 = 0,\label{m5}
\end{equation}
where $\overline{m}_1$ is the complex conjugate of $m_1$.\footnote{For type $D$
$w_{1}$ and $w_{2}$ are non-zero, whereas for type $O$ (or $N$ or $III$) 
both $w_{1}$ and $w_{2}$ vanish. In the set $CM$, for a general Ricci tensor, there remain 10 invariants
for type $D$, 9 for type $III$ (since $m_1R$=$m_2C$=$m_5R$=$0$ for the real ($R$) and imaginary ($C$) parts), 7 for
type $N$ (since $m_1C$=$m_2$=$m_5$=$0$) and of course 4 for type $O$.}  Equivalent
syzygies have been given by Zakhary and McIntosh for their set of
invariants \cite{ZM}. We observe the
following additional syzygies \cite{Musgrave}
\begin{equation}
6m_4 + w_1 r_2 = 0 \label{m4}
\end{equation}
and
\begin{equation}
m_3 - m_2 = 0. \label{m3}
\end{equation}
Further, we observe that \cite{santo} \footnote{Regarding syzygy (\ref{m1}) we note that $m_{1}=0 \Rightarrow 7r_{1}^{2}=12r_{3}$ and
that $7r_{1}^{2}=12r_{3} \Rightarrow m_{1}=0$ or $m_{1}w_{1}=3w_{2}r_{1}$.}
\begin{equation}
(-12r_{3}+7r_{1}^{2})w_{1}m_{1}-(12r_{2}^{2}-36r_{1}r_{3}+17r_{1}^{3})w_{2}=0 ,\label{m1}
\end{equation}
\begin{equation}
2(3m_6-m_1r_1)w_2+m_1^{2}w_1=0,\label{m6}
\end{equation}
and that
\begin{equation}
(-12r_{3}+7r_{1}^{2})^{3}-(12r_{2}^{2}-36r_{1}r_{3}+17r_{1}^{3})^{2}=0. \label{r3}
\end{equation}

The syzygy (\ref{r3}) is the counterpart for the  Plebanski tensor to the syzygy (\ref{elm})
for the Weyl tensor.\footnote{ We thank a referee for pointing this out.}
We expect that the syzygies (\ref{elm}) through (\ref{r3}) will hold in a wider class
of spacetimes than type $B$ warped products.

\section{The independent set of  invariants}
The syzygies (\ref{m2}) through (\ref{r3}) remove the independence of $m_2$,
$m_5$, $m_4$, $m_3$, $m_1$, $m_6$, and $r_3$ respectively.  According
to the syzygy (\ref{elm}) the sign of $w_2$ (which changes with signature)
cannot be obtained from $w_1$ whereas $w_1>0$. This suggests that for
metrics of the form (2) with (3) we take the set
\begin{equation}
\{R, r_1, r_2, w_2\} \label{MIN}
\end{equation}
as the independent set of scalar polynomial invariants, satisfying the
number of degrees of freedom in the curvature.

We note, however, that these do not necessarily constitute a
\emph{complete} set in the sense of classical invariant theory,
where the set is considered complete only if every invariant not
contained in the set can be written as an integral rational function
of the other members of the set \cite{Gurevich}. This is clearly not
the case here, for $w_1=\Psi_2^2$ can not be written as an integral
rational function of $w_2=\Psi_2^3$, nor vise versa, whereas $r_3$ is
determined as the root of a cubic from (\ref{r3}). 

From a practical standpoint, however, we are more concerned with
isolating the functional degrees of freedom in the curvature, since by
considering a more general interdependence (ie. by allowing
non-integer exponents in our polynomials) we can in principle solve
for any other of the remaining invariants in terms of the given
set. By inspection of the invariants listed in (\ref{MIN}), it is
clear that they are functionally independent and that they make up the
required number of degrees of freedom (\textit{i.e.} 4) for the class of spacetimes in
question, and in this sense we can consider them to be a `complete' list.

The invariants are reproduced in Appendix B.

\section{Some consequences of the syzygies}
The syzygies are certainly of value for computational efficiency.
However, understanding the physical benefit of the syzygies is
somewhat more difficult.  Mixed invariants have been used to provide a
number of alignment theorems for the perfect fluid and Maxwell cases
\cite{CM} and to classify the Riemann spinor \cite{Haddow}.  Our
purpose here is to provide some physical consequences of the syzygies
presented by way of expressing them in terms of the electric and
magnetic components of the Weyl tensor, and to show how the syzygy (\ref{r3})
can be used to impose restrictions on physical quantities in the energy-momentum tensor.

It is convenient to revert to classical
notation. First we consider the Weyl tensor ($C_{abcd}$). The
``electric" and ``magnetic" parts are defined, as usual, by
\begin{equation}
E_{ac}=C_{abcd}u^{b}u^{d},
\end{equation}
and
\begin{equation}
H_{a c}=C^{*~~~}_{abcd}u^{b}u^{d},
\end{equation}
where $u^{a}$ is a unit timelike vector and $C^*_{abcd} = \frac{1}{2}
\eta_{abef} {C^{ef}}_{cd}$ the dual tensor.  Writing
\begin{equation}
Q_{ab} = E_{ab} + iH_{ab}
\end{equation}
and
\begin{equation}
\overline{C}_{abcd} = C_{abcd} + iC^*_{abcd},
\end{equation}
directly from the expression
\begin{equation}
\label{exact}
4u_{[a} Q_{b][d}u_{c]} + g_{a[c}Q_{d]b} - g_{b[c}Q_{d]a} + i
 \eta_{abef} u^e u_{[c}{Q_{d]}}^f + i \eta_{cdef} u^e u_{[a}
 {Q_{b]}}^f
\end{equation}
for $-\frac{1}{2}\overline{C}_{abcd}$ (e.g., \cite{ES}) it follows that
\begin{equation}
\label{bonnor1}
\overline{C}_{abcd} \overline{C}^{abcd} = Q_{ab} Q^{ab}
\end{equation}
and that
\begin{equation}
\label{McI1}
{{C}_{ab}}^{cd} {{\overline{C}}_{ef}}^{ab} {{\overline{C}}_{cd}}^{ef} = Q^a_b Q^b_c Q^c_a .  
\end{equation} 
Equivalently, it follows that $w_{1}$ ($=2I$) is given by

\begin{equation}
\label{bonnor2}
\frac{1}{16}(E_{ab} E^{ab} - H_{ab} H^{ab}) + \frac{i}{8} (H_{ab}
E^{ab})
\end{equation}
and that $w_{2}$ ($=6J$) is given by
\begin{equation}
\label{McI2}
 \frac{1}{32} (3 E^a{}_b H^b{}_c H^c{}_a - E^a{}_b E^b{}_c E^c{}_a) 
	+ \frac{i}{32} (H^a{}_b H^b{}_c H^c{}_a - 3 E^a{}_b E^b{}_c
	H^c{}_a).
\end{equation}

Relations (\ref{bonnor2}) and (\ref{McI2}) hold \textit{in general} and are \textit{independent}
of the timelike vector chosen to split the Weyl tensor into its
electric and magnetic parts.  These expressions can, for example, be used to clarify
the relationship between the criterion of Bel and those of Misra and
Singh. The Bel condition \cite{Bel} is that $w_{1}$ = $w_{2}$ = 0 if
and only if the spacetime is of Petrov type N or III (or, of course
O). The condition of Misra and Singh \cite{MiS} can be stated as
$E_{ab} E^{ab} = H_{ab} H^{ab}$ and $E^a{}_b E^b{}_c E^c{}_a = H^a{}_b
H^b{}_c H^c{}_a$. Expressions (\ref{bonnor2}) and (\ref{McI2}) show
that these conditions are \textit{inequivalent}, contrary to the claim by
Zakharov \cite{Zak}.

Independence from the timelike vector chosen to split the Weyl tensor
does not, of course, extend into the mixed invariants. For example, it
follows that the real and imaginary parts of $8m_{1}$ are given by
\begin{equation}
C_{abcd} S^{ac} S^{bd} = 2 E^a_b S^b_c S^c_a - 2 E_{ab} S^{ab} S_{cd}
  u^c u^d + S^a_c u^c E_{ab} S^b_d u^d ,
\end{equation}
and 
\begin{equation}
{C}^*_{abcd} S^{ac} S^{bd} = 2 H^a_b S^b_c S^c_a - 2 H_{ab} S^{ab}
  S_{cd} u^c u^d + S^a_c u^c H_{ab} S^b_d u^d ,
\end{equation}
respectively, where $S_{ab}$ is the trace-free Ricci tensor. Expressions for the
next degree of mixed invariants ($m_2$ in the notation of \cite{CM}) have
been evaluated but the results are rather unwieldy (they contain 16
independent terms).  The important point is that with the appearance
of the vector $S^a_c u^c$ and scalar $S_{cd} u^c u^d$ it becomes
necessary to introduce an explicit form of $S_{ab}$ in order to proceed. 
This avenue is explored briefly below.

Relations equivalent to (\ref{bonnor1}) (or (\ref{bonnor2})) have been
reexamined recently by Bonnor \cite{Bonnor}.  Relations equivalent to
(\ref{McI1}) (or (\ref{McI2})) have been given by McIntosh {\it et
al.} \cite{McI} for Petrov types I (or D).

Specializing now to real $\Psi_{2}$ we observe that
\begin{equation}
\label{EH}
H_{ab} E^{ab} = 0 ,
\end{equation}
and that
\begin{equation}
\label{HHH}
H^a{}_b H^b{}_c H^c{}_a = 3 E^a{}_b E^b{}_c H^c{}_a .
\end{equation}
Moreover, from (9) we obtain the further relation
\begin{equation}
\label{EHH}
24 (3 E^a{}_b H^b{}_c H^c{}_a - E^a{}_b E^b{}_c E^c{}_a)^2 = 
	(E_{ab} E^{ab} - H_{ab} H^{ab})^3 .
\end{equation}
It is perhaps worth emphasizing the fact that relations (\ref{EH}),
(\ref{HHH}), and (\ref{EHH}) hold for all Petrov type $D$ spacetimes
with real $\Psi_{2}$. This includes a much wider class than class $B$ warped
product spaces.

We now wish to explore any restrictions that the syzygy (\ref{r3}) might impose on physical
quantities in the energy-momentum tensor. First, however, recall that at most three Ricci
invariants are independent. Finding syzygies associated with Ricci invariants of higher 
degree\footnote{These are constructed analogously to
$r_1$, $r_2$, and $r_3$ . For example,\\
$r_4 \equiv -\frac{1}{32} S_a{}^b S_b{}^c S_c{}^d S_d{}^e S_e{}^a 
\equiv  \Phi_{AB \dot{A}\dot{B}}\Phi^B{}_C{}^{\dot{B}}{}_{\dot{C}}\Phi^C{}_D{}^{\dot{C}}{}_{\dot{D}}
\Phi^D{}_E{}^{\dot{D}}{}_{\dot{E}}\Phi^{EA\dot{E}\dot{}}$.}
is a purely algorithmic procedure.\footnote{All possible combinations of $r_{1}$, $r_{2}$ and $r_{3}$
and $r_{n}$ (for each $n \geq 4$) are combined at degree $n+1$.  Expressing this
combination as a polynomial in the Ricci components, the coefficients
of any one of the Ricci components will form a linear system of
equations that can be solved to find the numerical coefficients
of the syzygy.} Ricci syzygies up to $r_{10}$ are given in Appendix C.

To begin, consider a perfect fluid,
\begin{equation}
T_{ab} = (\rho + p) u_a u_b + p g_{ab}
\end{equation}
where $\rho$ is the energy density, $p$ is the (isotropic) pressure, and
$u^a$ is the normalized timelike flow vector.  The Ricci invariants in this case all
reduce to the form
\begin{equation}
r_{n}=m_{n}(\rho + p)^{n+1},
\end{equation}
where $m_{n}$ is a constant. The syzygy (\ref{r3}) reduces to the identity $0 =
0$.  Increasing the complexity of the energy-momentum tensor
slightly does not offer anything new.  For both the case of
anisotropic pressure \cite{MM90}
\begin{equation}
T_{ab} = (\rho + p_2) u_a u_b + (p_1 - p_2) n_a n_b + p_2 g_{ab},
\end{equation}
where $n^a$ is a normalized spacelike vector orthogonal to $u^{a}$ (
$u_a n^a = 0$), and for the case of a perfect fluid with a heat flux
\cite{Ellis}, 
\begin{equation}
T_{ab} = (\rho + p) u_a u_b + p g_{ab} + 2 u_{(a}q_{b)},
\end{equation}
where $q^a$ is the heat flux vector such that $u_a q^a = 0$, 
the syzygy (\ref{r3}) again returns the identity $0 = 0$. No information
on possible relationships between the physical quantities that
describe the spacetime can be gained from the syzygy (\ref{r3}) in these cases.  
To obtain any restriction,  a more
complex physical decomposition must be considered.  The simplest
energy-momentum tensor for which the syzygy provides any
restriction is one with anisotropic pressure {\it and} a heat
flux:
\begin{equation}
T_{ab} = (\rho + p_2) u_a u_b + (p_1 - p_2) n_a n_b + p_2 g_{ab} + 2
u_{(a}q_{b)}.
\end{equation}
In this case the syzygy (\ref{r3}) reduces to
\begin{equation}
(q_{a}q^{a}-(n_{a}q^{a})^{2})^{2}(p1-p2)^{2}\textit{P}=0,\label{P}
\end{equation}
where $\textit{P}$ is a polynomial of degree 4 in 
\begin{equation}
\{\rho, p_1, p_2, n_a q^a \}
\end{equation}
and degree 3 in $q_aq^a$. If $p1\not=p2$ and $q_{a}q^{a}\not=(n_{a}q^{a})^{2}$
then the syzygy (\ref{r3}) reduces to the equation $\textit{P}=0$ and this 
reduces the number of independent physical scalars to four, the number of degrees of freedom. $\textit{P}$ is given by
\vspace{.25in}\\
$-p2^2\rho^4-p1^2\rho^4+2p1p2\rho^4-2p2^3\rho^3-4p2(n_{a}q^{a})^2\rho^3
+4(n_{a}q^{a})^2p1\rho^3+ \\ 2q_{a}q^{a}p2\rho^3 -2q_{a}q^{a}p1\rho^3-2p1^3\rho^3+2p1^2p2\rho^3
+2p1p2^2\rho^3-(q_{a}q^{a})^2\rho^2- \\ 2p1^3p2\rho^2+6(n_{a}q^{a})^2p1^2\rho^2-p1^4\rho^2 
+8q_{a}q^{a}p2^2\rho^2+2q_{a}q^{a}p1^2\rho^2-2p1p2^3\rho^2- \\ 6p2^2(n_{a}q^{a})^2\rho^2
-p2^4\rho^2-10q_{a}q^{a}p1p2\rho^2+6p1^2p2^2\rho^2+8(q_{a}q^{a})^2p1\rho- \\ 30p2^2(n_{a}q^{a})^2p1\rho
+2q_{a}q^{a}p2^3\rho+6p2^3(n_{a}q^{a})^2\rho+8q_{a}q^{a}p1^3\rho+2p1^3p2^2\rho 
+ \\ 30p1^2(n_{a}q^{a})^2p2\rho+2p2^3p1^2\rho+18q_{a}q^{a}p2(n_{a}q^{a})^2\rho-6p1^3(n_{a}q^{a})^2\rho
- \\ 2p1p2^4\rho+10q_{a}q^{a}p2^2p1\rho-10(q_{a}q^{a})^2p2\rho-2p2p1^4\rho-18q_{a}q^{a}(n_{a}q^{a})^2p1\rho
- \\ 20p1^2q_{a}q^{a}p2\rho+2p1^2p2^2q_{a}q^{a}-18q_{a}q^{a}p2^2(n_{a}q^{a})^2+10p1^3(n_{a}q^{a})^2p2
- \\ 10p2^3(n_{a}q^{a})^2p1-p2^4p1^2-(q_{a}q^{a})^2p2^2-4(n_{a}q^{a})^2p1^4-8p1^3q_{a}q^{a}p2+ \\ 4q_{a}q^{a}p1^4
-8p1(q_{a}q^{a})^2p2+4(q_{a}q^{a})^3+2p2^3p1q_{a}q^{a}+2p2^3p1^3+ \\ 54(n_{a}q^{a})^2p1q_{a}q^{a}p2
+4p2^4(n_{a}q^{a})^2-54(n_{a}q^{a})^4p1p2+27p2^2(n_{a}q^{a})^4+ \\ 27(n_{a}q^{a})^4p1^2-36q_{a}q^{a}(n_{a}q^{a})^2p1^2
-p1^4p2^2+8(q_{a}q^{a})^2p1^2$.

\newpage
It is a pleasure to thank Malcolm MacCallum for a number of useful
suggestions and two referees whose comments helped to improve this paper.
This work was supported in part by a grant (to KL) from
the Natural Sciences and Engineering Research Council of Canada.

\newpage

\section{Appendix A: Curvature components of the Class $B_1$ warped product spacetimes}
The following session demonstrates how \emph{GRTensorII} can be used to fix
the frame components of the curvature relative to a `standard' form by
applying $SL(2,C)$ rotations of the frame vectors. The
standard frame is defined so as to make use of this rotational freedom
in order to reduce the number of independent functions among the
curvature components. Once the frame is fixed as far as possible, the
subgroup of $SL(2,C)$ under which the resulting curvature
components are invariant is reported. The curvature components in the
standard frame are called `Karlhede invariants' for their usefulness
in determining the equivalence of spacetimes (e.g. \cite{McS}).

Within MapleV, we restart a fresh session and initialize the
\emph{GRTensorII} package.
\begin{maplegroup}
\begin{mapleinput}
\mapleinline{active}{1d}{restart:
}{%
}
\end{mapleinput}
\end{maplegroup}
\begin{maplegroup}
\begin{mapleinput}
\mapleinline{active}{1d}{grtw():}{%
}
\end{mapleinput}
\end{maplegroup}
\begin{maplegroup}
\begin{mapleinput}
\mapleinline{active}{1d}{qload (B1);}{%
}
\end{mapleinput}
\mapleresult
\begin{maplettyout}
Calculated ds for B1  
\end{maplettyout}
\begin{maplelatex}
\[
{\it Default\ spacetime}={\it B1}
\]
\end{maplelatex}

\begin{maplelatex}
\[
{\it Coordinates}
\]
\end{maplelatex}
\begin{maplelatex}
\[
{\it x\ }^{a}=[u, \,v, \,\theta , \,\phi ]
\]
\end{maplelatex}
\begin{maplelatex}
\[
{\it Line\ element}
\]
\end{maplelatex}
\begin{maplelatex}
\[
{\it \ ds}^{2}= - 2\,{\rm f}(u, \,v)\,{\it \ d}\,u^{\ }\,{\it d\ 
}\,v^{\ } + {\rm r}(u, \,v)^{2}\,{\rm g}(\theta , \,\phi )^{2}\,
{\it \ d}\,\theta ^{{\it 2\ }} + {\rm r}(u, \,v)^{2}\,{\rm g}(
\theta , \,\phi )^{2}\,{\it \ d}\,\phi ^{{\it 2\ }}
\]
\end{maplelatex}
\end{maplegroup}
We have loaded a line element corresponding to the class $B_1$ warped
product spacetimes. However, we will eventually wish to examine the
spinor components of the curvature. The function \texttt{nptetrad()}
can be used to \textit{automatically} create a null (specifically,
Newman-Penrose) tetrad from the metric. This tetrad will then be used
to calculate the curvature spinors in a spin frame corresponding to
the null frame.
\begin{maplegroup}
\begin{mapleinput}
\mapleinline{active}{1d}{nptetrad();}{%
}
\end{mapleinput}
\mapleresult
\begin{maplettyout}
The metric signature of the B1 spacetime is +2.
In order to create an NP-tetrad, the signature of g(dn,dn) will be
changed to -2.
Continue? (1=yes [default], other=no) :
\end{maplettyout}
\end{maplegroup}
\begin{maplegroup}
\begin{mapleinput}
\mapleinline{active}{1d}{1;
}{%
}
\end{mapleinput}
\mapleresult
\begin{maplettyout}
The signature of the B1 spacetime is now -2.
\end{maplettyout}
\begin{maplelatex}
\[
{\it For\ the\ B1\ spacetime:}
\]
\end{maplelatex}

\begin{maplelatex}
\[
{\it Basis\ inner\ product}
\]
\end{maplelatex}
\begin{maplelatex}
\[
\eta ^{{\it (a)}}\,^{{\it (b)}}= \left[ 
{\begin{array}{rrrr}
0 & 1 & 0 & 0 \\
1 & 0 & 0 & 0 \\
0 & 0 & 0 & -1 \\
0 & 0 & -1 & 0
\end{array}}
 \right] 
\]
\end{maplelatex}
\begin{maplelatex}
\[
{\it Null\ tetrad\ (covariant\ components)}
\]
\end{maplelatex}
\begin{maplelatex}
\[
{l_{a}}=[0, \,{\rm f}(u, \,v), \,0, \,0]
\]
\end{maplelatex}
\begin{maplelatex}
\[
{n_{a}}=[1, \,0, \,0, \,0]
\]
\end{maplelatex}
\begin{maplelatex}
\[
{m_{a}}= \left[  \! 0, \,0, \, - {\displaystyle \frac {1}{2}} \,I
\,\sqrt{2}\,{\rm g}(\theta , \,\phi )\,{\rm r}(u, \,v), \, - 
{\displaystyle \frac {1}{2}} \,\sqrt{2}\,{\rm g}(\theta , \,\phi 
)\,{\rm r}(u, \,v) \!  \right] 
\]
\end{maplelatex}
\begin{maplelatex}
\[
{{\it mbar}_{a}}= \left[  \! 0, \,0, \,{\displaystyle \frac {1}{2
}} \,I\,\sqrt{2}\,{\rm g}(\theta , \,\phi )\,{\rm r}(u, \,v), \,
 - {\displaystyle \frac {1}{2}} \,\sqrt{2}\,{\rm g}(\theta , \,
\phi )\,{\rm r}(u, \,v) \!  \right] 
\]
\end{maplelatex}
\begin{maplettyout}
The null tetrad has been stored as e(bdn,dn).
\end{maplettyout}
\end{maplegroup}
The curvature spinors are defined as the NP scalars \{\texttt{WeylSc},
\texttt{RicciSc}, and \texttt{Lambda}\} in the standard
\emph{GRTensorII} package. However, they are also defined
specifically as spinors \{\texttt{WeylSp}, \texttt{RicciSp}, and
\texttt{Lambda}\} in the \texttt{spinor.m} package, which we now
load. By calculating the latter objects, we can make use of the spinor
packages' facilities for performing frame rotations and fixing a
standard form for symmetric spinors.
\begin{maplegroup}
\begin{mapleinput}
\mapleinline{active}{1d}{grlib(spinor);}{%
}
\end{mapleinput}
\mapleresult
\begin{maplettyout}
Spinors and spacetime classification routines
Version 0.9 4 Dec 1997
\end{maplettyout}
\end{maplegroup}
\begin{maplegroup}
\begin{mapleinput}
\mapleinline{active}{1d}{grcalc(WeylSp,RicciSp,Lambda);}{%
}
\end{mapleinput}
\mapleresult
\begin{maplettyout}
Basis/tetrad related object definitions
Last modified 5 February 1997
\end{maplettyout}
\begin{maplettyout}
Calculated WeylSp for B1
\end{maplettyout}
\begin{maplettyout}
Calculated RicciSp for B1 
\end{maplettyout}
\begin{maplettyout}
Calculated Lambda for B1 
\end{maplettyout}

\end{maplegroup}

The next command expands the tensor components (the first argument,
`\texttt{\_}', tells the command to refer to the previously calculated
spinors) and applies a function which aliases the output to make it
more readable.

\begin{maplegroup}
\begin{mapleinput}
\mapleinline{active}{1d}{ gralter(_,expand,autoAlias);
}{%
}
\end{mapleinput}

\mapleresult
\begin{maplettyout}
Component simplification of a GRTensorII object:
 
Applying routine expand to object WeylSp
Applying routine expand to object RicciSp
Applying routine expand to object Lambda
Applying routine autoAlias to object WeylSp
Applying routine autoAlias to object RicciSp
Applying routine autoAlias to object Lambda
\end{maplettyout}

\end{maplegroup}
\begin{maplegroup}
\begin{mapleinput}
\mapleinline{active}{1d}{grdisplay(_);
}{%
}
\end{mapleinput}
\mapleresult
\begin{maplelatex}
\[
{\it For\ the\ B1\ spacetime:}
\]
\end{maplelatex}
\begin{maplelatex}
\begin{eqnarray*}
\lefteqn{{\Psi _{2}}\,{_{0}}=} \\
 & &  - {\displaystyle \frac {1}{6}} \,{\displaystyle \frac {{g_{
\phi }}^{2}}{r^{2}\,g^{4}}}  - {\displaystyle \frac {1}{6}} \,
{\displaystyle \frac {{g_{\theta }}^{2}}{r^{2}\,g^{4}}}  + 
{\displaystyle \frac {1}{3}} \,{\displaystyle \frac {{r_{u, \,v}}
}{r\,f}}  - {\displaystyle \frac {1}{3}} \,{\displaystyle \frac {
{r_{u}}\,{r_{v}}}{r^{2}\,f}}  - {\displaystyle \frac {1}{6}} \,
{\displaystyle \frac {{f_{u, \,v}}}{f^{2}}}  + {\displaystyle 
\frac {1}{6}} \,{\displaystyle \frac {{f_{u}}\,{f_{v}}}{f^{3}}} 
 + {\displaystyle \frac {1}{6}} \,{\displaystyle \frac {{g_{
\theta , \,\theta }}}{r^{2}\,g^{3}}}  + {\displaystyle \frac {1}{
6}} \,{\displaystyle \frac {{g_{\phi , \,\phi }}}{r^{2}\,g^{3}}} 
\end{eqnarray*}
\end{maplelatex}
\begin{maplelatex}
\[
{\Phi _{0}}\,{_{0}}={\displaystyle \frac {{r_{u}}\,{f_{u}}}{r\,f}
}  - {\displaystyle \frac {{r_{u, \,u}}}{r}} 
\]
\end{maplelatex}
\begin{maplelatex}
\[
{\Phi _{1}}\,{_{1}}={\displaystyle \frac {1}{2}} \,
{\displaystyle \frac {{r_{u}}\,{r_{v}}}{r^{2}\,f}}  + 
{\displaystyle \frac {1}{4}} \,{\displaystyle \frac {{g_{\phi }}
^{2}}{r^{2}\,g^{4}}}  + {\displaystyle \frac {1}{4}} \,
{\displaystyle \frac {{g_{\theta }}^{2}}{r^{2}\,g^{4}}}  - 
{\displaystyle \frac {1}{4}} \,{\displaystyle \frac {{f_{u, \,v}}
}{f^{2}}}  + {\displaystyle \frac {1}{4}} \,{\displaystyle 
\frac {{f_{u}}\,{f_{v}}}{f^{3}}}  - {\displaystyle \frac {1}{4}} 
\,{\displaystyle \frac {{g_{\theta , \,\theta }}}{r^{2}\,g^{3}}} 
 - {\displaystyle \frac {1}{4}} \,{\displaystyle \frac {{g_{\phi 
, \,\phi }}}{r^{2}\,g^{3}}} 
\]
\end{maplelatex}
\begin{maplelatex}
\[
{\Phi _{2}}\,{_{2}}= - {\displaystyle \frac {{r_{v, \,v}}}{r\,f^{
2}}}  + {\displaystyle \frac {{r_{v}}\,{f_{v}}}{r\,f^{3}}} 
\]
\end{maplelatex}
\begin{maplelatex}
\begin{eqnarray*}
\lefteqn{\Lambda ={\displaystyle \frac {1}{12}} \,{\displaystyle 
\frac {{g_{\phi }}^{2}}{r^{2}\,g^{4}}}  + {\displaystyle \frac {1
}{12}} \,{\displaystyle \frac {{g_{\theta }}^{2}}{r^{2}\,g^{4}}} 
 + {\displaystyle \frac {1}{6}} \,{\displaystyle \frac {{r_{u}}\,
{r_{v}}}{r^{2}\,f}}  + {\displaystyle \frac {1}{3}} \,
{\displaystyle \frac {{r_{u, \,v}}}{r\,f}}  + {\displaystyle 
\frac {1}{12}} \,{\displaystyle \frac {{f_{u, \,v}}}{f^{2}}}  - 
{\displaystyle \frac {1}{12}} \,{\displaystyle \frac {{f_{u}}\,{f
_{v}}}{f^{3}}}  - {\displaystyle \frac {1}{12}} \,{\displaystyle 
\frac {{g_{\theta , \,\theta }}}{r^{2}\,g^{3}}} } \\
 & & \mbox{} - {\displaystyle \frac {1}{12}} \,{\displaystyle 
\frac {{g_{\phi , \,\phi }}}{r^{2}\,g^{3}}} \mbox{\hspace{293pt}}
\end{eqnarray*}
\end{maplelatex}
\end{maplegroup}

At this point, we see that the Weyl spinor is already in what we would
regard as a `standard' form. That is, the $l$ and $n$ vectors of the
frame are aligned along the principle null directions so that the only
remaining Weyl component is $\Psi_2$. In this form, the Weyl spinor is
invariant under the $SL(2,C$) subgroups of spins and boosts, as
we can check using the function \texttt{isotest()}. Thus by requiring
the Weyl spinor to take this form, we have already lost four degrees
of frame rotational freedom.

\begin{maplegroup}
\begin{mapleinput}
\mapleinline{active}{1d}{isotest(WeylSp);
}{%
}
\end{mapleinput}
\mapleresult
\begin{maplelatex}
\[
\{{\it lnswap}, \,{\it Spin}, \,{\it Boost}\}
\]
\end{maplelatex}
\end{maplegroup}
(\texttt{lnswap} refers to the discrete group corresponding to a swap
of the $l$ and $n$ basis vectors). We can apply the same procedure to
the Ricci spinor,
\begin{maplegroup}
\begin{mapleinput}
\mapleinline{active}{1d}{isotest(RicciSp);}{%
}
\end{mapleinput}
\mapleresult
\begin{maplelatex}
\[
\{{\it lnswap}, \,{\it Spin}\}
\]
\end{maplelatex}
\end{maplegroup}
\noindent and learn that the Ricci spinor is, in fact, not invariant
under boosts of the frame. That is, we can remove this freedom from
the frame by using it to restrict the components of the Ricci tensor
which are non-invariant under boosts so that they fit some standard
form.  A set of standard forms for symmetric spinors is known to the
function \texttt{dytrgen()}, which can now be used to generate the
appropriate transformation of the frame. In this particular case, it
chooses a boost which will cause the $|\Phi_{00}|=|\Phi_{22}|$.
\begin{maplegroup}
\begin{mapleinput}
\mapleinline{active}{1d}{T:=dytrgen(Boost,RicciSp):}{%
}
\end{mapleinput}
\end{maplegroup}
\begin{maplegroup}
\begin{mapleinput}
\mapleinline{active}{1d}{evalm(T);
}{%
}
\end{mapleinput}
\mapleresult
\begin{maplelatex}
\[
 \left[ 
{\begin{array}{cc}
 \left(  \!  - {\displaystyle \frac { - {r_{v, \,v}}\,f + {r_{v}}
\,{f_{v}}}{f^{2}\,( - {r_{u}}\,{f_{u}} + {r_{u, \,u}}\,f)}}  \! 
 \right) ^{1/8} & 0 \\ [2ex]
0 & {\displaystyle \frac {1}{ \left(  \!  - {\displaystyle 
\frac { - {r_{v, \,v}}\,f + {r_{v}}\,{f_{v}}}{f^{2}\,( - {r_{u}}
\,{f_{u}} + {r_{u, \,u}}\,f)}}  \!  \right) ^{1/8}}} 
\end{array}}
 \right] 
\]
\end{maplelatex}
\end{maplegroup}
The dyad transformation $T$ is applied to the curvature spinors using
the \texttt{applydytr()} function. The following command creates a new
frame called \texttt{stdB} by applying the spinor dyad transformation,
\texttt{dytr=T}, to the curvature spinors. (In fact, since
\texttt{WeylSp} and \texttt{Lambda} are invariant under $T$, their
components are simply copied across to the new frame.)  By specifying
rotateTetrad, we also ensure that the frame vectors
$\{l,n,m,\overline{m}\}$ are themselves rotated.
\begin{maplegroup}
\begin{mapleinput}
\mapleinline{active}{1d}{applydytr(WeylSp,RicciSp,Lambda,dytr=T,\linebreak rotateTetrad=auto,newname=stdB1);}{%
}
\end{mapleinput}
\mapleresult
\begin{maplettyout}
Transformation completed.
The rotated frame has been named 'stdB1'.
\end{maplettyout}
\end{maplegroup}
\noindent We now switch to the new frame and examine the spinor
components.
\begin{maplegroup}
\begin{mapleinput}
\mapleinline{active}{1d}{grmetric(stdB1);
}{%
}
\end{mapleinput}
\mapleresult
\begin{maplettyout}
Default metric is now stdB1
\end{maplettyout}
\end{maplegroup}
\begin{maplegroup}
\begin{mapleinput}
\mapleinline{active}{1d}{ gralter(RicciSp,radsimp);
}{%
}
\end{mapleinput}
\mapleresult
\begin{maplettyout}
Component simplification of a GRTensorII object:
 Applying routine radsimp to object RicciSp
\end{maplettyout}

\end{maplegroup}
\begin{maplegroup}
\begin{mapleinput}
\mapleinline{active}{1d}{grdisplay(WeylSp,RicciSp,Lambda);
}{%
}
\end{mapleinput}
\mapleresult
\begin{maplelatex}
\[
{\it For\ the\ stdB1\ spacetime:}
\]
\end{maplelatex}
\begin{maplelatex}
\begin{eqnarray*}
\lefteqn{{\Psi _{2}}\,{_{0}}=} \\
 & &  - {\displaystyle \frac {1}{6}} \,{\displaystyle \frac {{g_{
\phi }}^{2}}{r^{2}\,g^{4}}}  - {\displaystyle \frac {1}{6}} \,
{\displaystyle \frac {{g_{\theta }}^{2}}{r^{2}\,g^{4}}}  + 
{\displaystyle \frac {1}{3}} \,{\displaystyle \frac {{r_{u, \,v}}
}{r\,f}}  - {\displaystyle \frac {1}{3}} \,{\displaystyle \frac {
{r_{u}}\,{r_{v}}}{r^{2}\,f}}  - {\displaystyle \frac {1}{6}} \,
{\displaystyle \frac {{f_{u, \,v}}}{f^{2}}}  + {\displaystyle 
\frac {1}{6}} \,{\displaystyle \frac {{f_{u}}\,{f_{v}}}{f^{3}}} 
 + {\displaystyle \frac {1}{6}} \,{\displaystyle \frac {{g_{
\theta , \,\theta }}}{r^{2}\,g^{3}}}  + {\displaystyle \frac {1}{
6}} \,{\displaystyle \frac {{g_{\phi , \,\phi }}}{r^{2}\,g^{3}}} 
\end{eqnarray*}
\end{maplelatex}
\begin{maplelatex}
\[
{\Phi _{0}}\,{_{0}}={\displaystyle \frac {I\,\sqrt{{r_{v, \,v}}\,
f - {r_{v}}\,{f_{v}}}\,\sqrt{{r_{u}}\,{f_{u}} - {r_{u, \,u}}\,f}
}{f^{2}\,r}} 
\]
\end{maplelatex}
\begin{maplelatex}
\begin{eqnarray*}
\lefteqn{{\Phi _{1}}\,{_{1}}=} \\{\displaystyle  \frac {1}{4}} ( 
 & & 2\,g^{4}\,{r_{u}}\,{r_{v}}\,f^{2} + f^{3}\,{g_{\phi }}^{2}
 + f^{3}\,{g_{\theta }}^{2} - g^{4}\,{f_{u, \,v}}\,r^{2}\,f + g^{
4}\,{f_{u}}\,{f_{v}}\,r^{2} - f^{3}\,g\,{g_{\theta , \,\theta }}
 - f^{3}\,g\,{g_{\phi , \,\phi }}) \\
 & &  \left/ {\vrule height0.37em width0em depth0.37em}
 \right. \!  \! (r^{2}\,g^{4}\,f^{3})
\end{eqnarray*}
\end{maplelatex}
\begin{maplelatex}
\[
{\Phi _{2}}\,{_{2}}={\displaystyle \frac {I\,\sqrt{{r_{v, \,v}}\,
f - {r_{v}}\,{f_{v}}}\,\sqrt{{r_{u}}\,{f_{u}} - {r_{u, \,u}}\,f}
}{f^{2}\,r}} 
\]
\end{maplelatex}
\begin{maplelatex}
\begin{eqnarray*}
\lefteqn{\Lambda ={\displaystyle \frac {1}{12}} \,{\displaystyle 
\frac {{g_{\phi }}^{2}}{r^{2}\,g^{4}}}  + {\displaystyle \frac {1
}{12}} \,{\displaystyle \frac {{g_{\theta }}^{2}}{r^{2}\,g^{4}}} 
 + {\displaystyle \frac {1}{6}} \,{\displaystyle \frac {{r_{u}}\,
{r_{v}}}{r^{2}\,f}}  + {\displaystyle \frac {1}{3}} \,
{\displaystyle \frac {{r_{u, \,v}}}{r\,f}}  + {\displaystyle 
\frac {1}{12}} \,{\displaystyle \frac {{f_{u, \,v}}}{f^{2}}}  - 
{\displaystyle \frac {1}{12}} \,{\displaystyle \frac {{f_{u}}\,{f
_{v}}}{f^{3}}}  - {\displaystyle \frac {1}{12}} \,{\displaystyle 
\frac {{g_{\theta , \,\theta }}}{r^{2}\,g^{3}}} } \\
 & & \mbox{} - {\displaystyle \frac {1}{12}} \,{\displaystyle 
\frac {{g_{\phi , \,\phi }}}{r^{2}\,g^{3}}} \mbox{\hspace{293pt}}
\end{eqnarray*}
\end{maplelatex}

\end{maplegroup}
\begin{maplegroup}

\end{maplegroup}
At this point the frame has been fixed as far as possible, as the
curvature spinor components are all invariant under the remaining
1-parameter continuous isotropy, spins. We can count the number of
independent functions among the components and subtract the
dimension of the frame freedom to determine that four scalar
polynomial invariants are expected.


\section{Appendix B: Invariants of Class $B$ warped product spacetimes}
 
We record here the invariants $\{R, r_1, r_2, w_2\}$. 

For type \textbf{$B_{1}$}:

\begin{maplelatex}
\[
{\it \ ds}^{2}= - 2\,{\rm f}(u, \,v)\,{\it \ d}\,u^{\ }\,{\it d\ 
}\,v^{\ } + {\rm r}(u, \,v)^{2}\,{\rm g}(\theta , \,\phi )^{2}\,
{\it \ d}\,\theta ^{{\it 2\ }} + {\rm r}(u, \,v)^{2}\,{\rm g}(
\theta , \,\phi )^{2}\,{\it \ d}\,\phi ^{{\it 2\ }}
\]
\end{maplelatex}
\begin{maplelatex}
\begin{eqnarray*}
\lefteqn{{\it R\ }=2( - {\rm r}(u, \,v)^{2}\,{\rm g}(\theta , \,
\phi )^{4}\,{f_{u}}\,{f_{v}} + {\rm r}(u, \,v)^{2}\,{\rm g}(
\theta , \,\phi )^{4}\,{f_{u, \,v}}\,{\rm f}(u, \,v)} \\
 & & \mbox{} + 4\,{\rm r}(u, \,v)\,{\rm g}(\theta , \,\phi )^{4}
\,{r_{u, \,v}}\,{\rm f}(u, \,v)^{2} + {\rm f}(u, \,v)^{3}\,{g_{
\phi }}^{2} - {\rm f}(u, \,v)^{3}\,{g_{\phi , \,\phi }}\,{\rm g}(
\theta , \,\phi ) \\
 & & \mbox{} + {\rm f}(u, \,v)^{3}\,{g_{\theta }}^{2} - {\rm f}(u
, \,v)^{3}\,{g_{\theta , \,\theta }}\,{\rm g}(\theta , \,\phi )
 + 2\,{\rm f}(u, \,v)^{2}\,{r_{u}}\,{\rm g}(\theta , \,\phi )^{4}
\,{r_{v}}) \left/ {\vrule height0.44em width0em depth0.44em}
 \right. \!  \! ({\rm f}(u, \,v)^{3} \\
 & & {\rm r}(u, \,v)^{2}\,{\rm g}(\theta , \,\phi )^{4})
\end{eqnarray*}
\end{maplelatex}

\begin{maplelatex}
\begin{eqnarray*}
\lefteqn{{\it R1\ }={\displaystyle \frac {1}{4}} (2\,{\rm f}(u, 
\,v)^{6}\,{g_{\phi }}^{2}\,{g_{\theta }}^{2} + {\rm f}(u, \,v)^{6
}\,{g_{\phi , \,\phi }}^{2}\,{\rm g}(\theta , \,\phi )^{2} + 
{\rm f}(u, \,v)^{6}\,{g_{\theta , \,\theta }}^{2}\,{\rm g}(\theta
 , \,\phi )^{2}} \\
 & & \mbox{} + 12\,{\rm r}(u, \,v)^{2}\,{\rm g}(\theta , \,\phi )
^{8}\,{f_{u}}\,{f_{v}}\,{\rm f}(u, \,v)^{2}\,{r_{u}}\,{r_{v}} + 
{\rm r}(u, \,v)^{4}\,{\rm g}(\theta , \,\phi )^{8}\,{f_{u, \,v}}
^{2}\,{\rm f}(u, \,v)^{2} \\
 & & \mbox{} + 2\,{\rm r}(u, \,v)^{2}\,{\rm g}(\theta , \,\phi )
^{5}\,{f_{u, \,v}}\,{\rm f}(u, \,v)^{4}\,{g_{\phi , \,\phi }} - 2
\,{\rm r}(u, \,v)^{2}\,{\rm g}(\theta , \,\phi )^{4}\,{f_{u, \,v}
}\,{\rm f}(u, \,v)^{4}\,{g_{\theta }}^{2} \\
 & & \mbox{} + 2\,{\rm r}(u, \,v)^{2}\,{\rm g}(\theta , \,\phi )
^{5}\,{f_{u, \,v}}\,{\rm f}(u, \,v)^{4}\,{g_{\theta , \,\theta }}
 - 4\,{\rm r}(u, \,v)^{2}\,{\rm g}(\theta , \,\phi )^{8}\,{f_{u, 
\,v}}\,{\rm f}(u, \,v)^{3}\,{r_{u}}\,{r_{v}} \\
 & & \mbox{} - 2\,{\rm r}(u, \,v)^{2}\,{\rm g}(\theta , \,\phi )
^{4}\,{f_{u, \,v}}\,{\rm f}(u, \,v)^{4}\,{g_{\phi }}^{2} + {\rm r
}(u, \,v)^{4}\,{\rm g}(\theta , \,\phi )^{8}\,{f_{u}}^{2}\,{f_{v}
}^{2} \\
 & & \mbox{} - 8\,{\rm f}(u, \,v)^{3}\,{\rm r}(u, \,v)^{2}\,{\rm 
g}(\theta , \,\phi )^{8}\,{r_{v, \,v}}\,{f_{u}}\,{r_{u}} - 8\,
{\rm f}(u, \,v)^{3}\,{\rm r}(u, \,v)^{2}\,{\rm g}(\theta , \,\phi
 )^{8}\,{f_{v}}\,{r_{v}}\,{r_{u, \,u}} \\
 & & \mbox{} - 2\,{\rm f}(u, \,v)^{6}\,{g_{\phi }}^{2}\,{g_{
\theta , \,\theta }}\,{\rm g}(\theta , \,\phi ) + 4\,{\rm f}(u, 
\,v)^{5}\,{g_{\phi }}^{2}\,{r_{u}}\,{\rm g}(\theta , \,\phi )^{4}
\,{r_{v}} \\
 & & \mbox{} - 2\,{\rm f}(u, \,v)^{6}\,{g_{\phi , \,\phi }}\,
{\rm g}(\theta , \,\phi )\,{g_{\theta }}^{2} + 2\,{\rm f}(u, \,v)
^{6}\,{g_{\phi , \,\phi }}\,{\rm g}(\theta , \,\phi )^{2}\,{g_{
\theta , \,\theta }} \\
 & & \mbox{} - 4\,{\rm f}(u, \,v)^{5}\,{g_{\phi , \,\phi }}\,
{\rm g}(\theta , \,\phi )^{5}\,{r_{u}}\,{r_{v}} - 2\,{\rm f}(u, 
\,v)^{6}\,{g_{\theta }}^{2}\,{g_{\theta , \,\theta }}\,{\rm g}(
\theta , \,\phi ) \\
 & & \mbox{} + 4\,{\rm f}(u, \,v)^{5}\,{g_{\theta }}^{2}\,{r_{u}}
\,{\rm g}(\theta , \,\phi )^{4}\,{r_{v}} - 2\,{\rm r}(u, \,v)^{4}
\,{\rm g}(\theta , \,\phi )^{8}\,{f_{u}}\,{f_{v}}\,{f_{u, \,v}}\,
{\rm f}(u, \,v) \\
 & & \mbox{} + 2\,{\rm r}(u, \,v)^{2}\,{\rm g}(\theta , \,\phi )
^{4}\,{f_{u}}\,{f_{v}}\,{\rm f}(u, \,v)^{3}\,{g_{\phi }}^{2} - 2
\,{\rm r}(u, \,v)^{2}\,{\rm g}(\theta , \,\phi )^{5}\,{f_{u}}\,{f
_{v}}\,{\rm f}(u, \,v)^{3}\,{g_{\phi , \,\phi }} \\
 & & \mbox{} + 2\,{\rm r}(u, \,v)^{2}\,{\rm g}(\theta , \,\phi )
^{4}\,{f_{u}}\,{f_{v}}\,{\rm f}(u, \,v)^{3}\,{g_{\theta }}^{2} - 
2\,{\rm r}(u, \,v)^{2}\,{\rm g}(\theta , \,\phi )^{5}\,{f_{u}}\,{
f_{v}}\,{\rm f}(u, \,v)^{3}\,{g_{\theta , \,\theta }} \\
 & & \mbox{} - 4\,{\rm f}(u, \,v)^{5}\,{g_{\theta , \,\theta }}\,
{\rm g}(\theta , \,\phi )^{5}\,{r_{u}}\,{r_{v}} + 4\,{\rm f}(u, 
\,v)^{4}\,{r_{u}}^{2}\,{\rm g}(\theta , \,\phi )^{8}\,{r_{v}}^{2}
 \\
 & & \mbox{} + 8\,{\rm f}(u, \,v)^{4}\,{\rm r}(u, \,v)^{2}\,{\rm 
g}(\theta , \,\phi )^{8}\,{r_{v, \,v}}\,{r_{u, \,u}} - 2\,{\rm f}
(u, \,v)^{6}\,{g_{\phi }}^{2}\,{g_{\phi , \,\phi }}\,{\rm g}(
\theta , \,\phi ) \\
 & & \mbox{} + {\rm f}(u, \,v)^{6}\,{g_{\phi }}^{4} + {\rm f}(u, 
\,v)^{6}\,{g_{\theta }}^{4}) \left/ {\vrule 
height0.44em width0em depth0.44em} \right. \!  \! ({\rm f}(u, \,v
)^{6}\,{\rm r}(u, \,v)^{4}\,{\rm g}(\theta , \,\phi )^{8})
\end{eqnarray*}
\end{maplelatex}

\begin{maplelatex}
\begin{eqnarray*}
\lefteqn{{\it R2\ }= - {\displaystyle \frac {3}{2}} ({\rm r}(u, 
\,v)^{2}\,{\rm g}(\theta , \,\phi )^{4}\,{f_{u}}\,{f_{v}} - {\rm 
r}(u, \,v)^{2}\,{\rm g}(\theta , \,\phi )^{4}\,{f_{u, \,v}}\,
{\rm f}(u, \,v) + {\rm f}(u, \,v)^{3}\,{g_{\phi }}^{2}} \\
 & & \mbox{} - {\rm f}(u, \,v)^{3}\,{g_{\phi , \,\phi }}\,{\rm g}
(\theta , \,\phi ) + {\rm f}(u, \,v)^{3}\,{g_{\theta }}^{2} - 
{\rm f}(u, \,v)^{3}\,{g_{\theta , \,\theta }}\,{\rm g}(\theta , 
\,\phi ) \\
 & & \mbox{} + 2\,{\rm f}(u, \,v)^{2}\,{r_{u}}\,{\rm g}(\theta , 
\,\phi )^{4}\,{r_{v}})( - {r_{v, \,v}}\,{\rm f}(u, \,v) + {f_{v}}
\,{r_{v}})\,({r_{u, \,u}}\,{\rm f}(u, \,v) - {f_{u}}\,{r_{u}})
 \left/ {\vrule height0.44em width0em depth0.44em} \right. \! 
 \! ( \\
 & & {\rm f}(u, \,v)^{7}\,{\rm r}(u, \,v)^{4}\,{\rm g}(\theta , 
\,\phi )^{4})\mbox{\hspace{227pt}}
\end{eqnarray*}
\end{maplelatex}

\begin{maplelatex}
\begin{eqnarray*}
\lefteqn{{\it W2R\ }= - {\displaystyle \frac {1}{36}} ({\rm r}(u
, \,v)^{2}\,{\rm g}(\theta , \,\phi )^{4}\,{f_{u, \,v}}\,{\rm f}(
u, \,v) - {\rm r}(u, \,v)^{2}\,{\rm g}(\theta , \,\phi )^{4}\,{f
_{u}}\,{f_{v}}} \\
 & & \mbox{} - 2\,{\rm r}(u, \,v)\,{\rm g}(\theta , \,\phi )^{4}
\,{r_{u, \,v}}\,{\rm f}(u, \,v)^{2} + {\rm f}(u, \,v)^{3}\,{g_{
\phi }}^{2} - {\rm f}(u, \,v)^{3}\,{g_{\phi , \,\phi }}\,{\rm g}(
\theta , \,\phi ) \\
 & & \mbox{} + {\rm f}(u, \,v)^{3}\,{g_{\theta }}^{2} - {\rm f}(u
, \,v)^{3}\,{g_{\theta , \,\theta }}\,{\rm g}(\theta , \,\phi )
 + 2\,{\rm f}(u, \,v)^{2}\,{r_{u}}\,{\rm g}(\theta , \,\phi )^{4}
\,{r_{v}})^{3} \left/ {\vrule height0.44em width0em depth0.44em}
 \right. \!  \! ({\rm f}(u, \,v)^{9} \\
 & & {\rm r}(u, \,v)^{6}\,{\rm g}(\theta , \,\phi )^{12})
\end{eqnarray*}
\end{maplelatex}
For type \textbf{$B_{2}$}:
\begin{maplelatex}
\[
{\it \ ds}^{2}={\rm f}(u, \,v)^{2}\,{\it \ d}\,u^{{\it 2\ }} + 
{\rm f}(u, \,v)^{2}\,{\it \ d}\,v^{{\it 2\ }} - 2\,{\rm r}(u, \,v
)^{2}\,{\rm g}(\theta , \,\phi )\,{\it \ d}\,\theta ^{\ }\,{\it d
\ }\,\phi ^{\ }
\]
\end{maplelatex}
\begin{maplelatex}
\begin{eqnarray*}
\lefteqn{{\it R\ }=2({\rm r}(u, \,v)^{2}\,{\rm g}(\theta , \,\phi
 )^{3}\,{f_{v}}^{2} - {\rm r}(u, \,v)^{2}\,{\rm g}(\theta , \,
\phi )^{3}\,{f_{v, \,v}}\,{\rm f}(u, \,v) + {\rm r}(u, \,v)^{2}\,
{\rm g}(\theta , \,\phi )^{3}\,{f_{u}}^{2}} \\
 & & \mbox{} - {\rm r}(u, \,v)^{2}\,{\rm g}(\theta , \,\phi )^{3}
\,{f_{u, \,u}}\,{\rm f}(u, \,v) - 2\,{\rm r}(u, \,v)\,{\rm g}(
\theta , \,\phi )^{3}\,{r_{u, \,u}}\,{\rm f}(u, \,v)^{2} \\
 & & \mbox{} - 2\,{\rm r}(u, \,v)\,{\rm g}(\theta , \,\phi )^{3}
\,{r_{v, \,v}}\,{\rm f}(u, \,v)^{2} - {\rm f}(u, \,v)^{4}\,{g_{
\theta }}\,{g_{\phi }} + {\rm f}(u, \,v)^{4}\,{g_{\phi , \,\theta
 }}\,{\rm g}(\theta , \,\phi ) \\
 & & \mbox{} - {\rm f}(u, \,v)^{2}\,{r_{u}}^{2}\,{\rm g}(\theta 
, \,\phi )^{3} - {\rm f}(u, \,v)^{2}\,{r_{v}}^{2}\,{\rm g}(\theta
 , \,\phi )^{3}) \left/ {\vrule height0.44em width0em depth0.44em
} \right. \!  \! ({\rm f}(u, \,v)^{4}\,{\rm r}(u, \,v)^{2}\,{\rm 
g}(\theta , \,\phi )^{3})\mbox{\hspace{16pt}}
\end{eqnarray*}
\end{maplelatex}
\begin{maplelatex}
\begin{eqnarray*}
\lefteqn{{\it R1\ }={\displaystyle \frac {1}{4}} ({\rm r}(u, \,v)
^{4}\,{\rm g}(\theta , \,\phi )^{6}\,{f_{v}}^{4} + {\rm r}(u, \,v
)^{4}\,{\rm g}(\theta , \,\phi )^{6}\,{f_{u}}^{4} + {\rm f}(u, \,
v)^{8}\,{g_{\theta }}^{2}\,{g_{\phi }}^{2}} \\
 & & \mbox{} + {\rm f}(u, \,v)^{8}\,{g_{\phi , \,\theta }}^{2}\,
{\rm g}(\theta , \,\phi )^{2} + {\rm f}(u, \,v)^{4}\,{r_{u}}^{4}
\,{\rm g}(\theta , \,\phi )^{6} + {\rm f}(u, \,v)^{4}\,{r_{v}}^{4
}\,{\rm g}(\theta , \,\phi )^{6} \\
 & & \mbox{} - 2\,{\rm r}(u, \,v)^{2}\,{\rm g}(\theta , \,\phi )
^{6}\,{f_{v, \,v}}\,{\rm f}(u, \,v)^{3}\,{r_{u}}^{2} - 2\,{\rm r}
(u, \,v)^{2}\,{\rm g}(\theta , \,\phi )^{6}\,{f_{v, \,v}}\,{\rm f
}(u, \,v)^{3}\,{r_{v}}^{2} \\
 & & \mbox{} - 2\,{\rm r}(u, \,v)^{4}\,{\rm g}(\theta , \,\phi )
^{6}\,{f_{u}}^{2}\,{f_{u, \,u}}\,{\rm f}(u, \,v) + 2\,{\rm r}(u, 
\,v)^{2}\,{\rm g}(\theta , \,\phi )^{3}\,{f_{u}}^{2}\,{\rm f}(u, 
\,v)^{4}\,{g_{\theta }}\,{g_{\phi }} \\
 & & \mbox{} - 2\,{\rm r}(u, \,v)^{2}\,{\rm g}(\theta , \,\phi )
^{4}\,{f_{u}}^{2}\,{\rm f}(u, \,v)^{4}\,{g_{\phi , \,\theta }} + 
10\,{\rm r}(u, \,v)^{2}\,{\rm g}(\theta , \,\phi )^{6}\,{f_{u}}^{
2}\,{\rm f}(u, \,v)^{2}\,{r_{u}}^{2} \\
 & & \mbox{} + 10\,{\rm r}(u, \,v)^{2}\,{\rm g}(\theta , \,\phi )
^{6}\,{f_{u}}^{2}\,{\rm f}(u, \,v)^{2}\,{r_{v}}^{2} + {\rm r}(u, 
\,v)^{4}\,{\rm g}(\theta , \,\phi )^{6}\,{f_{u, \,u}}^{2}\,{\rm f
}(u, \,v)^{2} \\
 & & \mbox{} - 2\,{\rm r}(u, \,v)^{2}\,{\rm g}(\theta , \,\phi )
^{3}\,{f_{u, \,u}}\,{\rm f}(u, \,v)^{5}\,{g_{\theta }}\,{g_{\phi 
}} + 2\,{\rm r}(u, \,v)^{2}\,{\rm g}(\theta , \,\phi )^{4}\,{f_{u
, \,u}}\,{\rm f}(u, \,v)^{5}\,{g_{\phi , \,\theta }} \\
 & & \mbox{} - 2\,{\rm r}(u, \,v)^{2}\,{\rm g}(\theta , \,\phi )
^{6}\,{f_{u, \,u}}\,{\rm f}(u, \,v)^{3}\,{r_{u}}^{2} - 2\,{\rm r}
(u, \,v)^{2}\,{\rm g}(\theta , \,\phi )^{6}\,{f_{u, \,u}}\,{\rm f
}(u, \,v)^{3}\,{r_{v}}^{2} \\
 & & \mbox{} + 2\,{\rm r}(u, \,v)^{2}\,{\rm g}(\theta , \,\phi )
^{6}\,{r_{u, \,u}}^{2}\,{\rm f}(u, \,v)^{4} - 8\,{\rm r}(u, \,v)
^{2}\,{\rm g}(\theta , \,\phi )^{6}\,{r_{u, \,u}}\,{\rm f}(u, \,v
)^{3}\,{f_{u}}\,{r_{u}} \\
 & & \mbox{} + 8\,{\rm r}(u, \,v)^{2}\,{\rm g}(\theta , \,\phi )
^{6}\,{r_{u, \,u}}\,{\rm f}(u, \,v)^{3}\,{r_{v}}\,{f_{v}} - 4\,
{\rm r}(u, \,v)^{2}\,{\rm g}(\theta , \,\phi )^{6}\,{r_{u, \,u}}
\,{\rm f}(u, \,v)^{4}\,{r_{v, \,v}} \\
 & & \mbox{} + 8\,{\rm r}(u, \,v)^{2}\,{\rm g}(\theta , \,\phi )
^{6}\,{f_{u}}\,{r_{u}}\,{\rm f}(u, \,v)^{3}\,{r_{v, \,v}} + 2\,
{\rm r}(u, \,v)^{4}\,{\rm g}(\theta , \,\phi )^{6}\,{f_{v}}^{2}\,
{f_{u}}^{2} \\
 & & \mbox{} - 2\,{\rm r}(u, \,v)^{4}\,{\rm g}(\theta , \,\phi )
^{6}\,{f_{v}}^{2}\,{f_{u, \,u}}\,{\rm f}(u, \,v) + 2\,{\rm r}(u, 
\,v)^{2}\,{\rm g}(\theta , \,\phi )^{3}\,{f_{v}}^{2}\,{\rm f}(u, 
\,v)^{4}\,{g_{\theta }}\,{g_{\phi }} \\
 & & \mbox{} - 2\,{\rm r}(u, \,v)^{2}\,{\rm g}(\theta , \,\phi )
^{4}\,{f_{v}}^{2}\,{\rm f}(u, \,v)^{4}\,{g_{\phi , \,\theta }} + 
10\,{\rm r}(u, \,v)^{2}\,{\rm g}(\theta , \,\phi )^{6}\,{f_{v}}^{
2}\,{\rm f}(u, \,v)^{2}\,{r_{u}}^{2} \\
 & & \mbox{} + 10\,{\rm r}(u, \,v)^{2}\,{\rm g}(\theta , \,\phi )
^{6}\,{f_{v}}^{2}\,{\rm f}(u, \,v)^{2}\,{r_{v}}^{2} - 16\,{\rm f}
(u, \,v)^{3}\,{\rm r}(u, \,v)^{2}\,{\rm g}(\theta , \,\phi )^{6}
\,{r_{u, \,v}}\,{f_{v}}\,{r_{u}} \\
 & & \mbox{} - 16\,{\rm f}(u, \,v)^{3}\,{\rm r}(u, \,v)^{2}\,
{\rm g}(\theta , \,\phi )^{6}\,{r_{u, \,v}}\,{f_{u}}\,{r_{v}} - 2
\,{\rm f}(u, \,v)^{8}\,{g_{\theta }}\,{g_{\phi }}\,{g_{\phi , \,
\theta }}\,{\rm g}(\theta , \,\phi ) \\
 & & \mbox{} + 2\,{\rm f}(u, \,v)^{6}\,{g_{\theta }}\,{g_{\phi }}
\,{r_{u}}^{2}\,{\rm g}(\theta , \,\phi )^{3} + 2\,{\rm f}(u, \,v)
^{6}\,{g_{\theta }}\,{g_{\phi }}\,{r_{v}}^{2}\,{\rm g}(\theta , 
\,\phi )^{3} \\
 & & \mbox{} - 2\,{\rm f}(u, \,v)^{6}\,{g_{\phi , \,\theta }}\,
{\rm g}(\theta , \,\phi )^{4}\,{r_{u}}^{2} - 2\,{\rm f}(u, \,v)^{
6}\,{g_{\phi , \,\theta }}\,{\rm g}(\theta , \,\phi )^{4}\,{r_{v}
}^{2} \\
 & & \mbox{} + 2\,{\rm f}(u, \,v)^{4}\,{r_{u}}^{2}\,{\rm g}(
\theta , \,\phi )^{6}\,{r_{v}}^{2} + 8\,{\rm f}(u, \,v)^{4}\,
{\rm r}(u, \,v)^{2}\,{\rm g}(\theta , \,\phi )^{6}\,{r_{u, \,v}}
^{2} \\
 & & \mbox{} - 2\,{\rm r}(u, \,v)^{4}\,{\rm g}(\theta , \,\phi )
^{6}\,{f_{v}}^{2}\,{f_{v, \,v}}\,{\rm f}(u, \,v) + {\rm r}(u, \,v
)^{4}\,{\rm g}(\theta , \,\phi )^{6}\,{f_{v, \,v}}^{2}\,{\rm f}(u
, \,v)^{2} \\
 & & \mbox{} + 2\,{\rm r}(u, \,v)^{4}\,{\rm g}(\theta , \,\phi )
^{6}\,{f_{v, \,v}}\,{\rm f}(u, \,v)^{2}\,{f_{u, \,u}} - 2\,{\rm r
}(u, \,v)^{2}\,{\rm g}(\theta , \,\phi )^{3}\,{f_{v, \,v}}\,{\rm 
f}(u, \,v)^{5}\,{g_{\theta }}\,{g_{\phi }} \\
 & & \mbox{} + 2\,{\rm r}(u, \,v)^{2}\,{\rm g}(\theta , \,\phi )
^{4}\,{f_{v, \,v}}\,{\rm f}(u, \,v)^{5}\,{g_{\phi , \,\theta }}
 + 2\,{\rm r}(u, \,v)^{2}\,{\rm g}(\theta , \,\phi )^{6}\,{r_{v, 
\,v}}^{2}\,{\rm f}(u, \,v)^{4} \\
 & & \mbox{} - 8\,{\rm r}(u, \,v)^{2}\,{\rm g}(\theta , \,\phi )
^{6}\,{r_{v}}\,{f_{v}}\,{\rm f}(u, \,v)^{3}\,{r_{v, \,v}} - 2\,
{\rm r}(u, \,v)^{4}\,{\rm g}(\theta , \,\phi )^{6}\,{f_{v, \,v}}
\,{\rm f}(u, \,v)\,{f_{u}}^{2}) \left/ {\vrule 
height0.44em width0em depth0.44em} \right. \!  \! ( \\
 & & {\rm f}(u, \,v)^{8}\,{\rm r}(u, \,v)^{4}\,{\rm g}(\theta , 
\,\phi )^{6})
\end{eqnarray*}
\end{maplelatex}
\begin{maplelatex}
\begin{eqnarray*}
\lefteqn{{\it R2\ }= - {\displaystyle \frac {3}{8}} (4\,{\rm f}(u
, \,v)\,{f_{u}}\,{r_{u}}\,{r_{v, \,v}} + 4\,{f_{v}}^{2}\,{r_{v}}
^{2} - 4\,{\rm f}(u, \,v)\,{r_{u, \,u}}\,{f_{u}}\,{r_{u}}} \\
 & & \mbox{} - 2\,{r_{u, \,u}}\,{\rm f}(u, \,v)^{2}\,{r_{v, \,v}}
 + {r_{u, \,u}}^{2}\,{\rm f}(u, \,v)^{2} + 4\,{f_{u}}^{2}\,{r_{u}
}^{2} + {r_{v, \,v}}^{2}\,{\rm f}(u, \,v)^{2} + 4\,{f_{u}}^{2}\,{
r_{v}}^{2} \\
 & & \mbox{} + 4\,{f_{v}}^{2}\,{r_{u}}^{2} + 4\,{r_{u, \,v}}^{2}
\,{\rm f}(u, \,v)^{2} + 4\,{r_{u, \,u}}\,{\rm f}(u, \,v)\,{r_{v}}
\,{f_{v}} - 8\,{\rm f}(u, \,v)\,{r_{u, \,v}}\,{f_{u}}\,{r_{v}} \\
 & & \mbox{} - 4\,{r_{v}}\,{f_{v}}\,{\rm f}(u, \,v)\,{r_{v, \,v}}
 - 8\,{r_{u, \,v}}\,{\rm f}(u, \,v)\,{f_{v}}\,{r_{u}})({\rm f}(u
, \,v)^{4}\,{g_{\theta }}\,{g_{\phi }} - {\rm f}(u, \,v)^{4}\,{g
_{\phi , \,\theta }}\,{\rm g}(\theta , \,\phi ) \\
 & & \mbox{} + {\rm f}(u, \,v)^{2}\,{r_{u}}^{2}\,{\rm g}(\theta 
, \,\phi )^{3} + {\rm f}(u, \,v)^{2}\,{r_{v}}^{2}\,{\rm g}(\theta
 , \,\phi )^{3} + {\rm r}(u, \,v)^{2}\,{\rm g}(\theta , \,\phi )
^{3}\,{f_{v}}^{2} \\
 & & \mbox{} - {\rm r}(u, \,v)^{2}\,{\rm g}(\theta , \,\phi )^{3}
\,{f_{v, \,v}}\,{\rm f}(u, \,v) + {\rm r}(u, \,v)^{2}\,{\rm g}(
\theta , \,\phi )^{3}\,{f_{u}}^{2} \\
 & & \mbox{} - {\rm r}(u, \,v)^{2}\,{\rm g}(\theta , \,\phi )^{3}
\,{f_{u, \,u}}\,{\rm f}(u, \,v)) \left/ {\vrule 
height0.44em width0em depth0.44em} \right. \!  \! ({\rm f}(u, \,v
)^{10}\,{\rm r}(u, \,v)^{4}\,{\rm g}(\theta , \,\phi )^{3})
\end{eqnarray*}
\end{maplelatex}
\begin{maplelatex}
\begin{eqnarray*}
\lefteqn{{\it W2R\ }= - {\displaystyle \frac {1}{36}} ({\rm r}(u
, \,v)^{2}\,{\rm g}(\theta , \,\phi )^{3}\,{f_{v}}^{2} - {\rm r}(
u, \,v)^{2}\,{\rm g}(\theta , \,\phi )^{3}\,{f_{v, \,v}}\,{\rm f}
(u, \,v)} \\
 & & \mbox{} + {\rm r}(u, \,v)^{2}\,{\rm g}(\theta , \,\phi )^{3}
\,{f_{u}}^{2} - {\rm r}(u, \,v)^{2}\,{\rm g}(\theta , \,\phi )^{3
}\,{f_{u, \,u}}\,{\rm f}(u, \,v) \\
 & & \mbox{} + {\rm r}(u, \,v)\,{\rm g}(\theta , \,\phi )^{3}\,{r
_{v, \,v}}\,{\rm f}(u, \,v)^{2} + {\rm r}(u, \,v)\,{\rm g}(\theta
 , \,\phi )^{3}\,{r_{u, \,u}}\,{\rm f}(u, \,v)^{2} - {\rm f}(u, 
\,v)^{4}\,{g_{\theta }}\,{g_{\phi }} \\
 & & \mbox{} + {\rm f}(u, \,v)^{4}\,{g_{\phi , \,\theta }}\,{\rm 
g}(\theta , \,\phi ) - {\rm f}(u, \,v)^{2}\,{r_{u}}^{2}\,{\rm g}(
\theta , \,\phi )^{3} - {\rm f}(u, \,v)^{2}\,{r_{v}}^{2}\,{\rm g}
(\theta , \,\phi )^{3})^{3} \left/ {\vrule 
height0.44em width0em depth0.44em} \right. \!  \! ( \\
 & & {\rm f}(u, \,v)^{12}\,{\rm r}(u, \,v)^{6}\,{\rm g}(\theta , 
\,\phi )^{9})
\end{eqnarray*}
\end{maplelatex}

\section{Appendix C: Ricci syzygies}
Syzygies for the Ricci invariants (in terms of $r_{1}$, $r_{2}$ and $r_{3}$) up to $r_{10}$ are given by:

\begin{equation}
\label{eqn-syr4}
6r_4 - 5 r_1 r_2 = 0 ,
\end{equation}
\begin{equation}
\label{eqn-syr5}
24r_{5}+3r_{1}^3-8r_{2}^2-18r_{1}r_{3}=0 ,
\end{equation}
\begin{equation}
\label{eqn-syr6}
24 r_6 -7 r_2 r_1{}^2 -14 r_3 r_2 = 0, 
\end{equation}
\begin{equation}
\label{eqn-syr7}
144r_{7}-36r_{3}^2-64r_{1}r_{2}^2-36r_{1}^2r_{3}+9r_{1}^4=0, 
\end{equation}
\begin{equation}
\label{eqn-syr8}
36r_{8}-4r_{2}^3-27r_{1}r_{2}r_{3}=0, 
\end{equation}
\begin{equation}
\label{eqn-syr9}
576r_{9}-160r_{2}^{2}r_{3}-180r_{1}r_{3}^{2}-160r_{1}^{2}r_{2}^{2}+9r_{1}^{5}=0, 
\end{equation}
and
\begin{equation}
\label{eqn-syr10}
1728r_{10}-396r_{2}r_{3}^{2}-352r_{1}r_{2}^{3}-792r_{1}^{2}r_{2}r_{3}+99r_{1}^{4}r_{2}=0.
\end{equation}
These syzygies hold \textit{in general}, unlike the syzygy (\ref{r3}).
\end{document}